\documentclass[preprint2]{proto}
\usepackage{times}

\begin{document}
\title{\textbf{\LARGE Disk-Planet Interactions During Planet Formation }}
\author {\textbf{\large J. C. B. Papaloizou}}
\affil{{\it University of Cambridge}}
\author {\textbf{\large R. P. Nelson}}
\affil{\small\em {\it Queen Mary, University of London}}
\author {\textbf{\large W. Kley}}
\affil{{\it Universit${\ddot {\rm a}}$t T${\ddot {\rm u}}$bingen}}
\author {\textbf{\large F. S. Masset}}
\affil{{\it Saclay, France and UNAM Mexico }}
\author {\textbf{\large  P. Artymowicz}}
\affil{{\it University of Toronto at Scarborough and University of Stockholm}}

\begin{abstract}

\noindent The discovery of close orbiting extrasolar giant planets
led to extensive studies of disk planet interactions and the forms of migration
that can result as a means of accounting for their location.
Early work established the type I and type II migration regimes for
low mass embedded planets and high mass gap forming planets respectively.
While providing an attractive means of accounting for close orbiting planets
intially formed at several $AU,$ inward migration 
times for objects in the earth mass range
were found to be disturbingly short, making the survival of giant planet cores
an issue.
Recent progress 
in this area has come from the application of modern numerical techniques wich make use of
up to date supercomputer resources. These have enabled higher resolution
studies of the regions close to the planet and the initiation of studies of planets interacting
with  disks undergoing MHD turbulence. This  work has led to indications of how the  inward migration of low to intermediate mass planets could be slowed down or reversed. 
 In addition,  the possibility of a 
  new very fast type~III migration regime, that can be directed
inwards or outwards, that is relevant to 
partial gap forming planets  in massive disks  has been investigated.

\vspace{1cm}
\end{abstract}

\section{INTRODUCTION}
The discovery of extrasolar planets around sun--like
 stars (51~Pegasi~b) ({\it  Mayor and Queloz}, 1995; {\it Marcy and Butler}, 1995; 
  {\it Marcy and Butler}, 1998)
 has revealed a population of close orbiting giant planets
 with periods of typically a few days, the so called 'hot Jupiters'.
 The difficulties associated with forming such planets in situ, either in the critical 
 core mass accumulation followed by gas accretion scenario, or the gravitational 
 instability scenario for giant planet formation has led  to the realisation of the
 potential importance of large scale migration in forming or young planetary systems.
 
 This in turn led to more intensive theoretical development  of disk protoplanet interaction theory
 that had already led to predictions of orbital migration (see {\it  Lin and Papaloizou}, 1993;
{\it Lin et al.}, 2000 and references therein). At the time of PPIV, the type~I and type~II
 migration regimes, the former applying to small mass embedded protoplanets
 and the latter to gap forming massive protoplanets, had become apparent.
 Both these regimes predicted disturbingly short radial infall times
 that in the type~I case threatened the survival of embryo cores in the $1-15 M_{\oplus}$
 regime before they could accrete gas to become giant planets. The main questions 
 to be addressed were how to resolve the type~I migration issue and to confirm that
 type~II migration applicable to giant planets could indeed account for the observed 
 radial distribution and the hot Jupiters.

 Here, we review recent progress in the field of disk planet interactions in the context
 of orbital migration. For reasons of space constraint we shall not consider
 the problem of excitation or damping  of orbital eccentricity.
  The most recent progress 
 in this area has come from
 carrying out large scale two and  three dimension simulations that require the most
 up to date supercomputer resources. This has enabled the study of disk planet interactions
 in  disks undergoing MHD turbulence, the study of the regions close to the planet
 using high resolution multigrid techniques, led to suggestions for the  possible resolution of the type~I issue
 and revealed another possible type~III migration regime. However, the complex nature of these
 problems makes them challenging numerically and  as a consequence numerical convergence
 has not been attained in some cases.
 
 In sections~{\ref{sec:typeI}, \ref{sec:typeII}, and \ref{sec:typeIII}
  we review type~I migration, type~II migration and type~III migration respectively.
  In section~\ref{sec:turb} we review recent work on disk planet interactions in disks
  with MHD turbulence and in section~\ref{sec:Disc} we give a summary.

\section{TYPE I MIGRATION}
\label{sec:typeI}
When the mass of the protoplanet is small
the response it induces in the disk can be calculated 
using linear theory. When the disk flow is non magnetic and laminar,
density waves propagate both outwards and inwards away from the protoplanet.
These waves carry positive and negative angular momentum respectively and
accordingly a compensating tidal torque is applied to the orbit resulting
in type~I migration. 
\subsection{The tidal torque}
The problem of determining  the evolution
of the planet orbit amounts to an evaluation of tidal
torques.  For a sufficiently small planet mass (an upper limit of which
will be specified below) one supposes that the gravitational
potential of the protoplanet forces small perturbations.
The hydrodynamic equations are then linearized about
a basic state consisting of an unperturbed axisymmetric accretion disk
and the response calculated.
The gravitational potential
$\psi$ of a proplanet in circular orbit is expressed as  a Fourier series in
the form
\begin{equation}
\psi(r,\varphi,t)=\sum_{m=0}^\infty\psi_m(r)\cos\{m[\varphi-\omega_p t ]\},
\end{equation}
where $\varphi$ is the azimuthal angle and $2\pi /(\omega_p )$ is the orbital period of the planet
of mass $M_p$ at orbital semi-major axis $a.$
The total torque acting on the disk is given by
$\Gamma = - \int_{Disk} \Sigma \vec{r} \times \nabla \psi d^2 r$ 
where $\Sigma$ is the surface density of the disk.

An  external forcing potential $\psi_m(r,\varphi)$ with azimuthal mode number $m$ 
that rotates with a pattern frequency $\omega_p$ in a disk with angular
velocity $\Omega(r)$ triggers a response that exchanges angular momentum with
the orbit whenever, neglecting effects due to pressure, 
$m(\Omega-\omega_p)$ is equal either $0$ or $\pm\kappa,$ with, for a Keplerian disk to adequate acuracy,
$\kappa \equiv \Omega$ being the epicyclic frequency. The first possibility occurs when $\Omega =\omega_p$ and 
thus corresponds to a co-rotation resonance. The second
possibility corresponds to an inner Lindblad resonance located inside the orbit for
$\Omega=\omega_p +\kappa /m $ and an outer Lindblad resonance outside the orbit for $\Omega= \omega_p -\kappa /m .$

\noindent {\em 2.1.1. Torques at Lindblad resonances}.
Density waves are launched at Lindblad resonances and as a consequence 
of this a torque acts on the planet.
It is possible to solve the wave excitation problem using the WKB method.
In that approximation an analytic expression for the torque can be found.
The torque arising  from the component of the potential with azimuthal
mode number $m$ is found, for a Keplerian disk, to be given by
\begin{equation}
\Gamma^{\rm LR}_m = {{{\rm sign}(\omega_p-\Omega)\pi^2\Sigma}\over
{3\Omega\omega_p}} \Psi^2, \label{eqn:lindbladone}
\end{equation}
with
\begin{equation}
\Psi = r\frac{d\psi_m}{dr}+\frac{2m^2(\Omega-\omega_p)}{\Omega}\psi_m .
\end{equation}
where the expresion has to be evaluated at the location of
the resonance.

The derivation of this torque formula in the context
of satellite(planet)- interaction  with a gaseous disk can be found in ( {\it Goldreich and Tremaine}, 1979; {\it Lin and Papaloizou}, 1979; {\it Lin and Papaloizou}, 1993). 
In a Keplerian disk, the torque exerted on the planet
from an outer Lindblad resonance is negative corresponding to a drag, and
the torque due to  an inner Lindblad resonance is positive corresponding to an acceleration.

The total torque may be obtained by summing contributions over $m.$
However, when doing so it must be borne in mind that the above analysis,
appropriate to a cold disk, is only valid 
when 
$\xi = mc_s/(r\Omega) <<1,$

For finite $\xi$ the positions of the Lindblad resonances
are modified, being now given by
\begin{equation}
\label{eqn:disprel}
 m^2(\Omega-\omega_p)^2=\Omega^2(1+ \xi^2),
\end{equation}
where $c_s$ is the sound speed. 
 
The  effective positions of the resonances are shifted with respect to
the cold disk case . In particular, noting that $c_s = H\Omega,$ with $H <<r,$
being the disk semithickness one sees that when $m\rightarrow\infty$, Lindblad
resonances pile up at
\begin{equation}
r=a \pm {2 H\over 3}.
\end{equation}
Physically these locations correspond to where the disk flow relative to the planet becomes
sonic so that they are naturally the points from where the density waves are first launched  
({\it Goodman and Rafikov}, 2001).

In addition a correction factor has to be applied to the expression for the torque acting on the disk
which now reads (see {\it Artymowicz}, 1993; {\it Ward}, 1997; {\it Papaloizou and Larwood}, 2000)
\begin{equation}
\label{eqn:lindbladtwo}
\Gamma^{\rm LR}_m = {{{\rm sign}(\omega_p-\Omega)\pi^2\Sigma }\over
{3\Omega\omega_p\sqrt{1+\xi^2}(1+4\xi^2)}} \Psi^2.
\end{equation}

This together with the shift in Lindblad resonance locations ensures
that when contributions are summed over $m,$ they decrease rapidly for $\xi >>1,$
a phenomenon known as the torque cut off.

\noindent{\em 2.1.2. Differential Lindblad torque}.
The total outer  (resp. inner)  Lindblad  torque are obtained by summing
over all individual components.
\begin{equation}
\Gamma_{\rm OLR (ILR)} =\sum_{m=1(2)}^{+\infty}\Gamma_m^{\rm OLR (ILR)},
\end{equation}

These are
referred to  as one-sided Lindblad torques.  They scale with
$h^{-3}$, where $h=H/r$ is the disk aspect ratio
({\it Ward}, 1997).

\begin{figure}[ht]
\centering \includegraphics[width=0.80\linewidth]{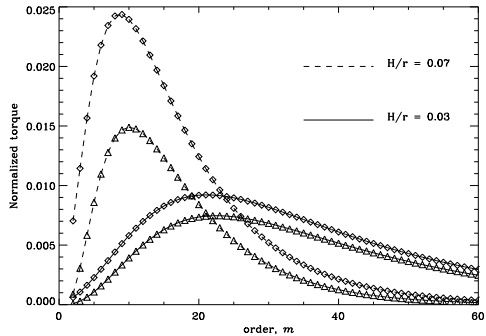}
\caption{\label{fig:torques}Individual inner and outer torques 
(in absolute value) in a
$h=0.07$ and $h=0.03$ disk, as a function of $m$. For each
disk thickness, the upper curve (diamonds) shows the outer torque
and the lower one (triangles) the inner torque.
These torques are normalized to $\Gamma_0= \pi q^2\Sigma a^4\omega_p^2h^{-3}$.}
\end{figure}

In Fig.~\ref{fig:torques} one can see that the torque
cut-off occurs at larger $m$ values in the thinner disk (the outer
torque value peaks around $m\sim 8-9$ for $h=0.07$, while it peaks
around $m\sim 21-22$ for $h=0.03$).  Also, there is for both disk
aspect ratios a very apparent mismatch between the inner and the outer
torques, the former being systematically smaller than the later. If we
consider the torque of the disk acting on the planet, then the outer
torques are negative and the inner ones positive, and the total torque
is therefore negative. As a consequence {\it migration is directed inwards}
and leads to a decay of the orbit onto the central object
({\it Ward}, 1997).
It can be shown that the relative mismatch of inner and outer torques
scales with the disk thickness ({\it Ward}, 1997).
Since one-sided torques scale as
$h^{-3}$, the migration rate scales with $h^{-2}$.

There are several  reasons for the torque asymmetry  which conspire to
make  the  differential Lindblad  torque  a  sizable  fraction of  the
one-sided torque in a $h=O(10^{-1})$ disk ({\it Ward}, 1997).
Most importantly,  for a  given $m$ value,  the inner  Lindblad resonances
lie  further from  the  orbit than  the  corresponding outer  Lindlad
resonances. From this it follows that the relative disk motion
becomes sonic further away inside the orbit making the launching of 
density waves less efficient in the inner regions.
Note that the net torque on the disk is positive making that on the planet
negative so producing inward migration.
This is found to be the case for disks with reasonable density profiles
(see Eq. (\ref{eq:tanaka}) below.

\noindent{ \em 2.1.3. Linear corotation torque }.\label{CRTO}
The angular momentum exchange  at a corotation resonance  corresponds
to different physical processes than at a Lindblad resonance.  At the
latter the perturbing potential tends to excite epicyclic motion, and,
in a protoplanetary disk, the  angular momentum deposited is evacuated
through pressure supported waves.  On  the contrary, theses waves  are
evanescent in the  corotation  region, and  are unable  to remove the
angular momentum brought by the perturber ({\it Goldreich and Tremaine}, 1979).

The corotation  torque  exerted  on  a  disk by  an   external
perturbing  potential with $m$ fold symmetry is given by
\begin{equation}
\label{eqn:corone}
\Gamma^{\rm CR}_{\rm m}=\frac{m\pi^2\psi^2}{2(d\Omega/dr)}
\frac{d}{dr}\left(\frac{\Sigma}{B}\right),
\end{equation}
to be  evaluated at the corotation  radius.
$\psi$ is the amplitude of the forcing potential, and $B=\kappa^2/(4\Omega)$
the second  Oort's constant.
Since  $B$ is  half the  flow vorticity,
the corotation torque  scales with  the gradient  of (the
inverse  of)  the  specific   vorticity,  sometimes  also  called  the
vortensity.   The  corotation  torque   therefore  cancels  out  in  a
$\Sigma\propto r^{-3/2}$  disk, such as the standard minimum  mass solar nebula with $h=0.05$
(MMSN).

In most cases, a disk sharp edge being
a possible  exception,  the  corotation torque can be
safely neglected when estimating  the
migration timescale in the linear regime. Indeed, even the fully
unsaturated corotation torque amounts at most to a few tens of
percent of the differential Lindblad torque
({\it Ward}, 1997; {\it Tanaka et al.}, 2002),
while {\it Korycansky and Pollack} (1993) find
through numerical integrations that the corotation torque is an even
smaller fraction of the differential Lindblad torque than given by
analytical estimates.

\noindent{\em 2.1.4. Nonlinear effects}.
In a frame which corotates with the perturbation pattern,
if inviscid, the  flow  in the neighborhood of  corotation   consists of libration
islands, in  which fluid elements librate on  closed streamlines,
between regions in which fluid elements circulate in opposite senses.
This is true for general perturbations and not just those
with $m$ fold symmetry. See, e.g., fig.~\ref{fig:horseshoe} which applies to the corotation or coorbital
region associated with a planet in circular orbit.
In linear theory, the period of libration, which tends to infinity
as the  perturbation amplitude (or planet mass)  tends to zero,  is  such that
complete libration cycles do not occur and angular momentum exchange rates are 
appropriate only to sections at definite radial locations.

However, in actual fact, fluid elements in the libration region exchange  zero net angular momentum
with the perturbation during one complete libration cycle.
Accordingly if a steady state of this type  can be set up, in full non linear theory the  net
corotation torque is zero. When this is the case the corotation resonance 
is said to be saturated.

But note that, when present, viscosity can cause an exchange 
of angular momentum between  librating  and circulating fluid elements  
 which results in a net corotation  torque.
Then saturation is prevented.
This is possible if the viscous timescale across the libration
islands is smaller than the libration time (see 
{\it Goldreich and Sari}, 2003; {\it Ogilvie and Lubow}, 2003).
It is found that for small viscosity, the corotation torque is proportional to
$\nu$ ({\it Balmforth and Korycansky}, 2001), while at large viscosity one obtains
the torque induced as material flowing through the orbit passes by the perturbing
planet ({\it Masset}, 2001).

Note  that these saturation  properties  are  not
captured  by a  linear analysis,  since saturation  requires  a finite
libration time, hence  a finite resonance width. In  the linear limit,
the corotation torque appears as  a discontinuity at corotation of the
advected  angular  momentum  flux,  which  corresponds  to  infinitely
narrow, fully unsaturated libration islands.

\begin{figure}
\centering
\includegraphics[width=0.80\linewidth]{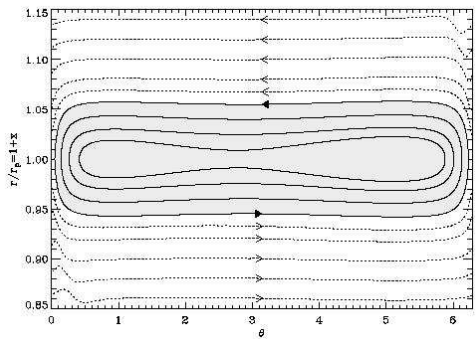}
\caption{\label{fig:horseshoe}Horseshoe region with closed
streamlines as viewed in a frame corotating with
 a planet in circular orbit (shaded area). The planet is located at
$r=1$ and $\theta=0$ or $2\pi$.  }
\end{figure}

\begin{figure}[ht]
\centering
\includegraphics[width=0.90\linewidth]{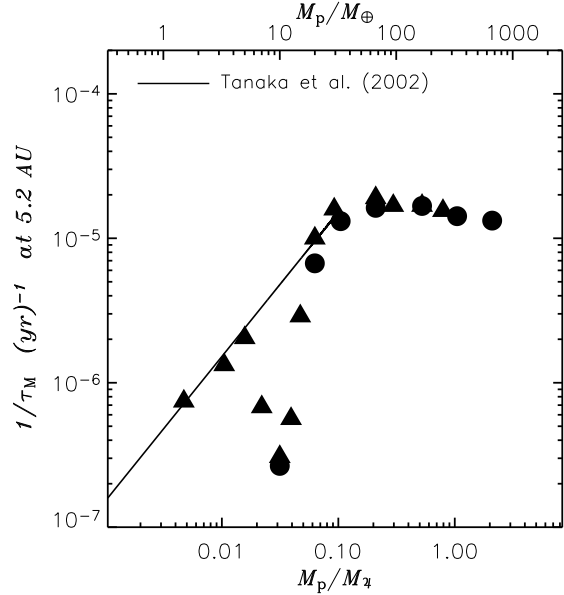}
\caption{The  migration rate for different planet
masses for 3D fully non-linear nested grid simulations.
The symbols denote different approximations (smoothening) for
the potential of the planet. The solid line refers to linear results
for Type I migration by {\it Tanaka et al.}, (2002),
see Eq.~\ref{eq:tanaka}.
Figure adapted from {\it D'Angelo et al.} (2003).
\label{fig:mig1}}
\end{figure}
\subsection{Type I migration drift rate estimates}
\label{subsec:migratI}
There are a number of  estimates of the Type I migration rate  in the
literature that are based on
summing resonant torque contributions
(see {\it Ward}, 1997; {\it Papaloizou and  Larwood}, 2000 and references  therein).
Calculations using 2D models have to soften the planet potential to obtain agreement with results
obtained from  3D calculations but such agreement may be obtained for reasonable
choices of the softening parameter (see {\it Papaloizou and Terquem}, 2006).
 
The most  recent linear calculations by  {\it Tanaka et al.} (2002) that  take into
account 3D  effects, and are based upon the value of the total
tidal  torque,  including the corotation torque (fully  unsaturated
since it is a linear estimate),  give
\begin{equation}
\label{eq:tanaka}
\tau \, \equiv \, a/\dot{a} \, = \,
 (2.7+1.1\alpha)^{-1}\frac{M_*^2}{M_p\Sigma a^2}h^2\omega_p^{-1},
\end{equation}
for a surface density profile $\Sigma \propto r^{-\alpha}$.
For an Earth-mass planet around a solar mass star at $r=1$~AU, in a
disk with $\Sigma=1700$~g~cm$^{-2}$ and $h=0.05$, this translates
into $\tau=1.6\cdot 10^5$~years.

This semi-analytic estimate has been verified by means of 3D numerical
simulations ({\it Bate et al.}, 2003; {\it D'Angelo et al.}, 2003a).  Both
find an excellent agreement in the limit of low-mass, thus they
essentially validate the linear analytical estimate.  However,
while {\it Bate et al.} (2003) find agreement with
the linear results for all planet masses,
{\it D'Angelo et al.} (2003a) find very long migration rates
for intermediate masses, i.e. for Neptune-sized objects (see
Fig.~\ref{fig:mig1}). 
Additional 2d and 3D high resolution numerical simulations by 
{\it Masset et al.} (2006) show that this {\it migration offset} from the
linear results is a robust phenomenon whose strength
varies {\it i)} with departure from the
$\Sigma \propto r^{-3/2}$ relation, {\it ii)} with the value
of the viscosity, and {\it iii)} with the disk thickness. 
The transition from linear to the offset regime is apparently caused by 
the onset of non-linear effects that could be related to 
corotation torques whose strength also increase with departure
from  $\Sigma \propto r^{-3/2}$ ( see section~2.1.3). 

The type I migration time scale is very short, much shorter than the
build up time of the $M_p\sim 5-15$~$M_\oplus$ solid core of a giant
planet (see, e.g., {\it Papaloizou and Nelson}, 2005 for a discussion).
Hence, the existence of type I migration makes potential difficulties for the
accumulation scenario for these massive cores. This 
remains a problem within the framework of 
planet formation theory (see the discussion below for possible resolutions).  

The influence of the disk's self-gravity on type I migration 
has been analyzed through numerical and semi-analytical
methods ({\it Nelson and Benz}, 2003; {\it Pierens and  Hur{\'e}}, 2003). 
It increases the migration rate but the effect is small
for typical disk parameters.

{\it Papaloizou and Larwood} (2000) incorporate a non zero planet eccentricity in their torque
calculations and find that in general the torques weaken with increasing eccentricity
and can even reverse once the eccentricity exceeds $h$ by some factor of order unity.
Thus a process that maintains eccentricity could potentially help to stall the migration
process. A similar effect occurs if the disk has a global non axisymmetric distortion
such as occurs if it has a finite eccentricity. This can also  result in a weakening
of the tidal interaction and a stalling of the torques under appropriate conditions
(see {\it Papaloizou}, 2002).
 
Recent attempts to include  more
detailed physics of the protoplanetary disk, such as opacity
transitions and their impact on the disk profile ({\it Menou and Goodman}, 2004)
, or radiative transfer and the importance
of shadowing in the planet vicinity ({\it Jang-Condell and Sasselov}, 2005 ), have
lead to lower estimates of the type I migration rates which might help
resolve the accretion/migration timescale discrepancy.
The aforementioned offset and also effects due to magnetic fields (see, e.g., {\it Terquem}, 2003)
and their associated turbulence (see below) may
help to extend the type I migration time scale  and allow proto giant
planet cores to form.
\begin{figure}[t]
\centering
\includegraphics[width=0.80\linewidth]{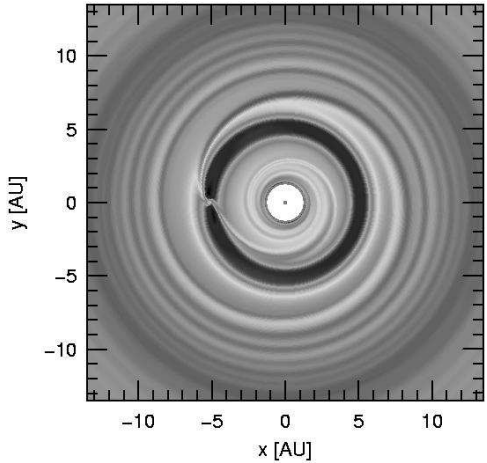}
\caption{Surface density profile for an initially axisymmetric
disk model two hundred orbital orbits after the introduction  of a planet
that subsequently remains on a fixed circular.
The mass ratio is $10^{-3}.$
\label{fig:pm3-200}}
\end{figure}
\begin{figure*}[ht]
\resizebox{1.00\textwidth}{!}{%
\includegraphics[clip]{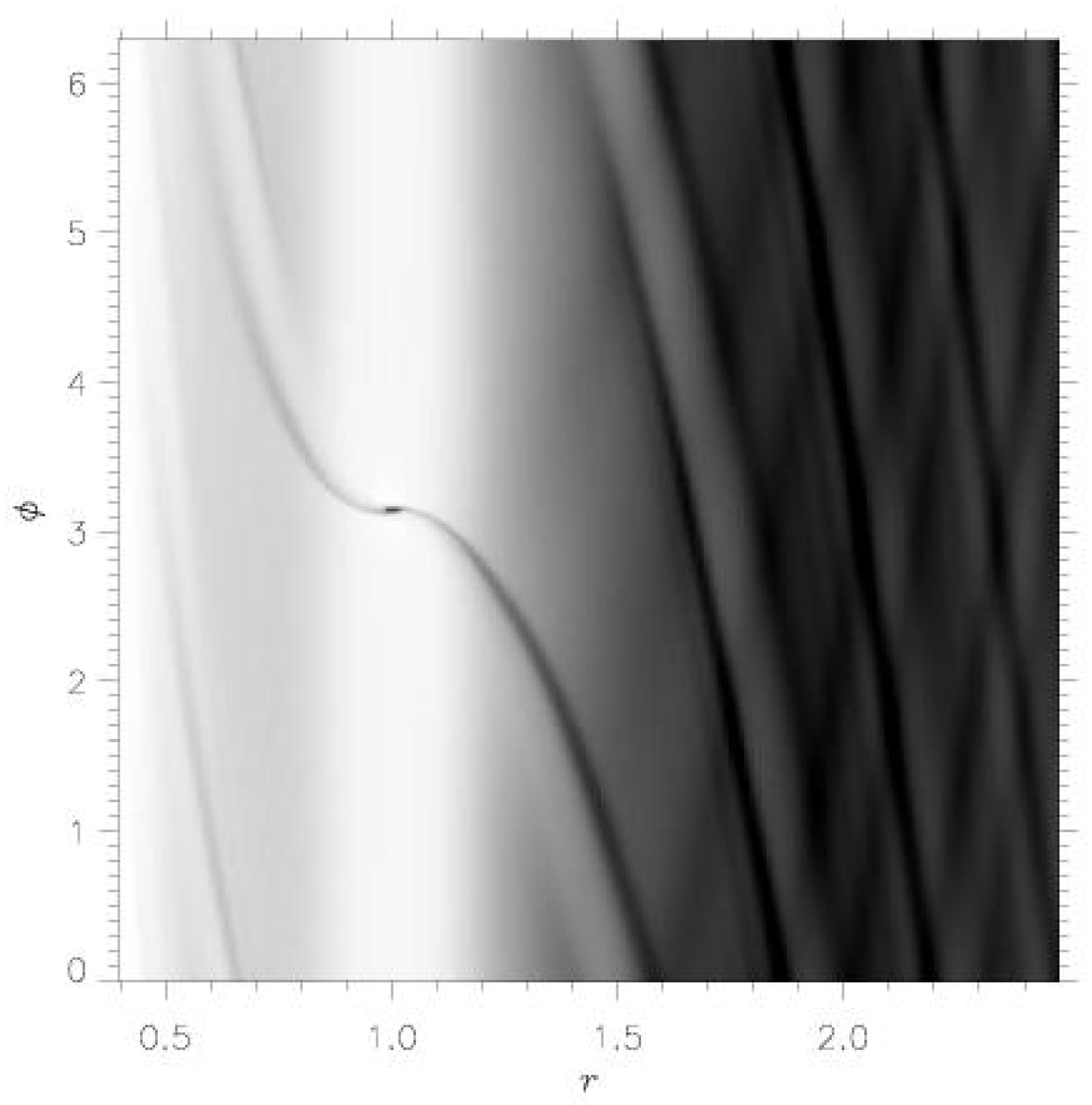}%
\includegraphics[clip]{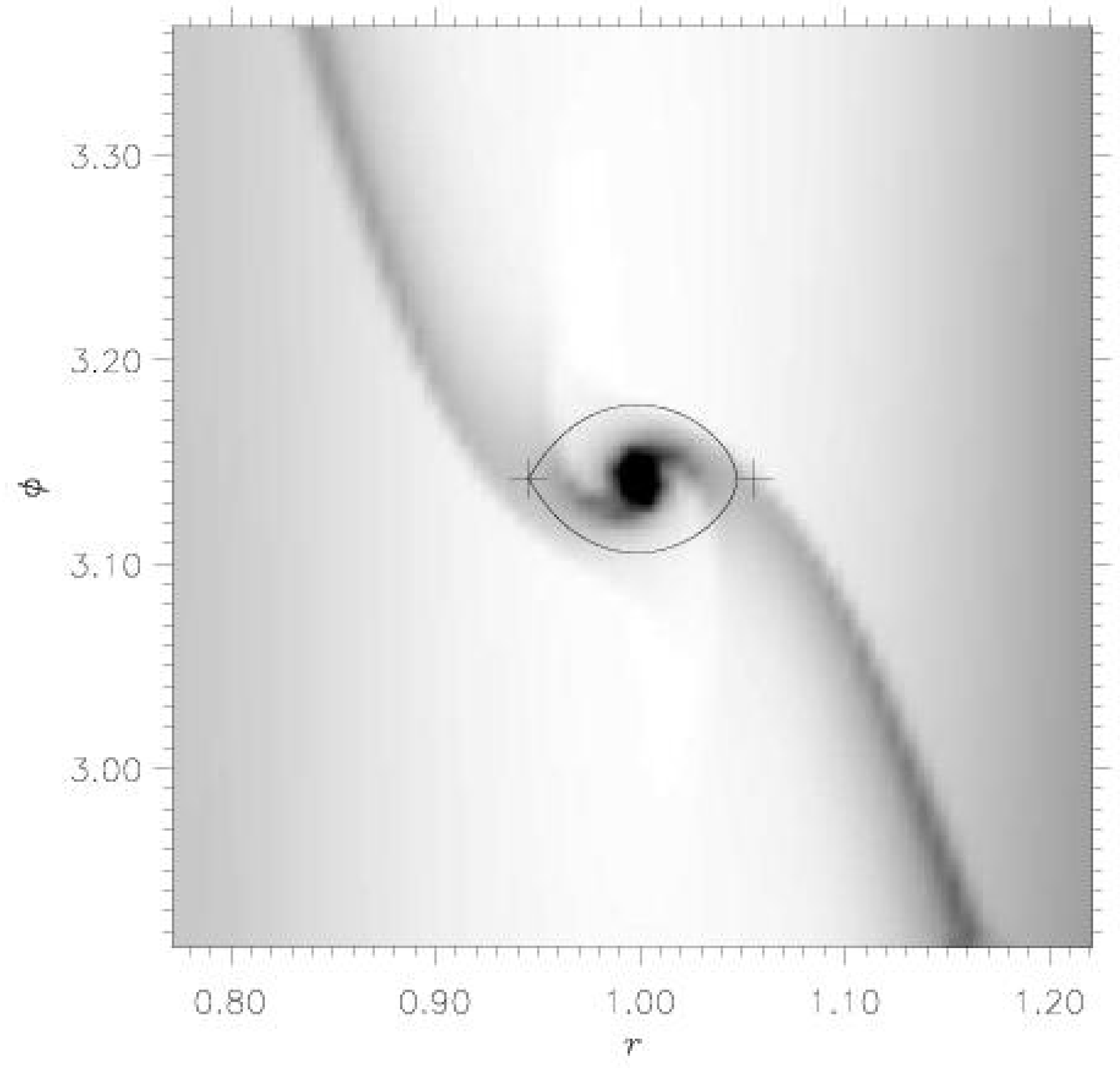}%
\includegraphics[clip]{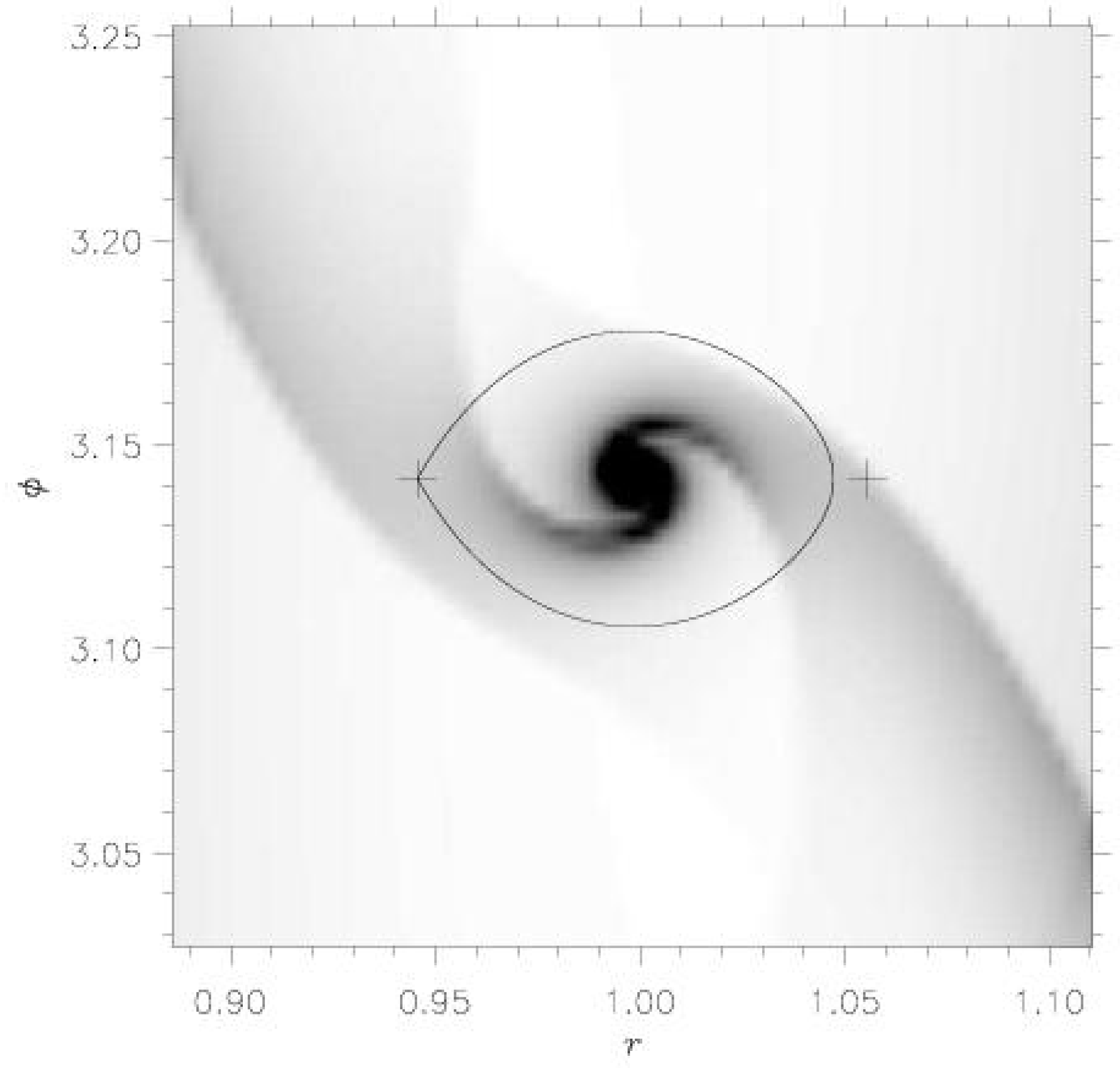}}\\
\resizebox{1.00\textwidth}{!}{%
\includegraphics[clip]{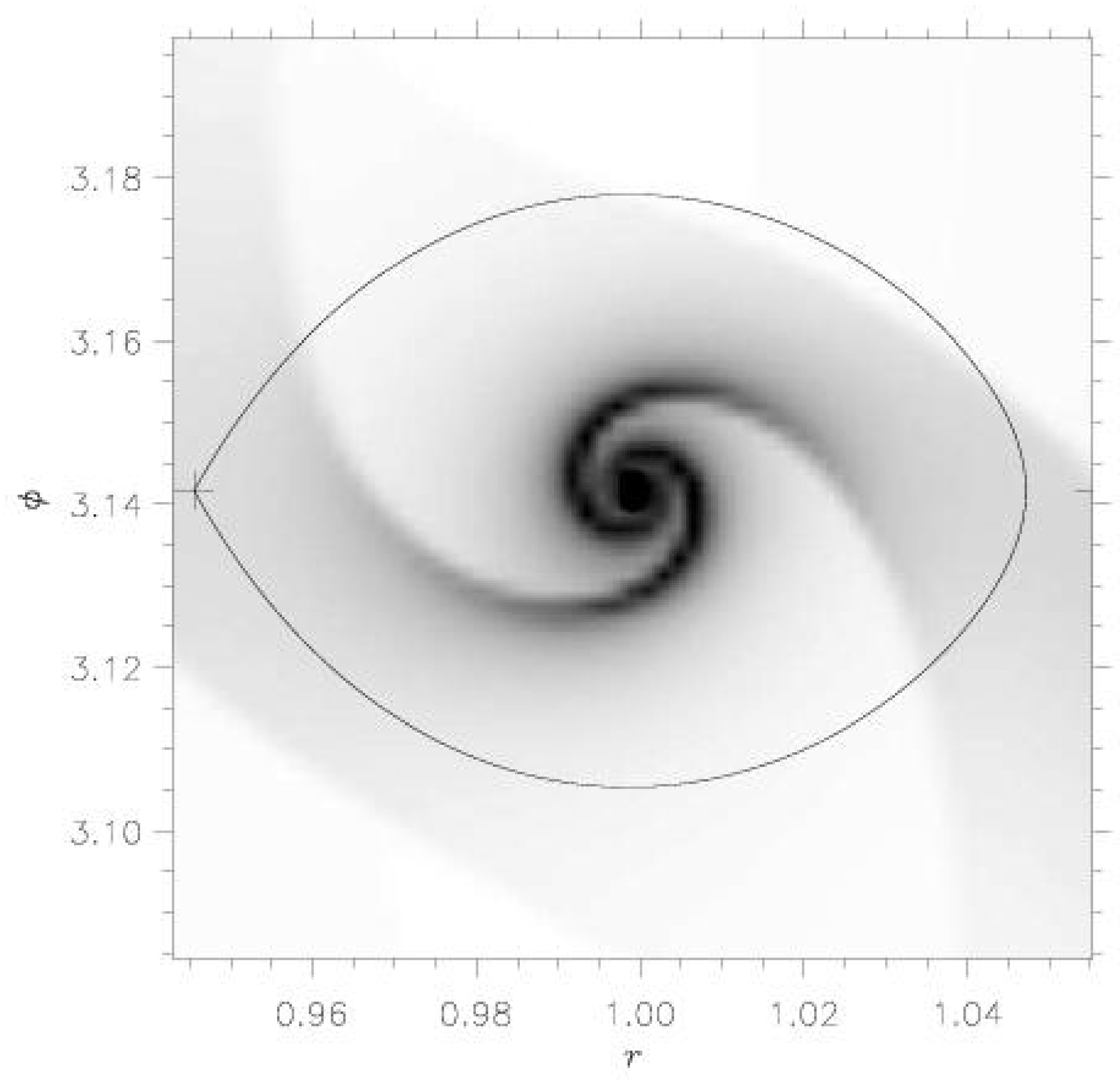}%
\includegraphics[clip]{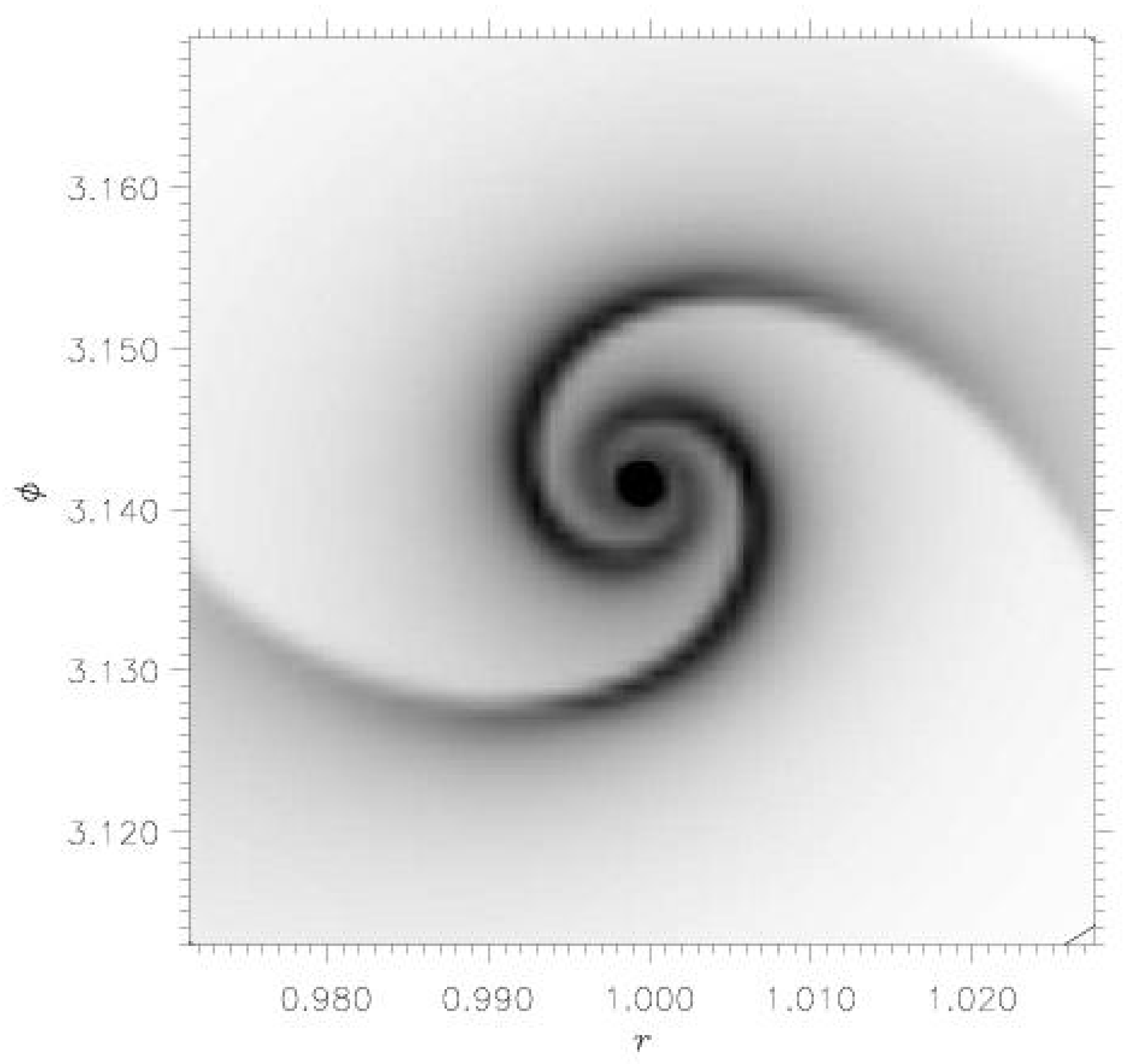}%
\includegraphics[clip]{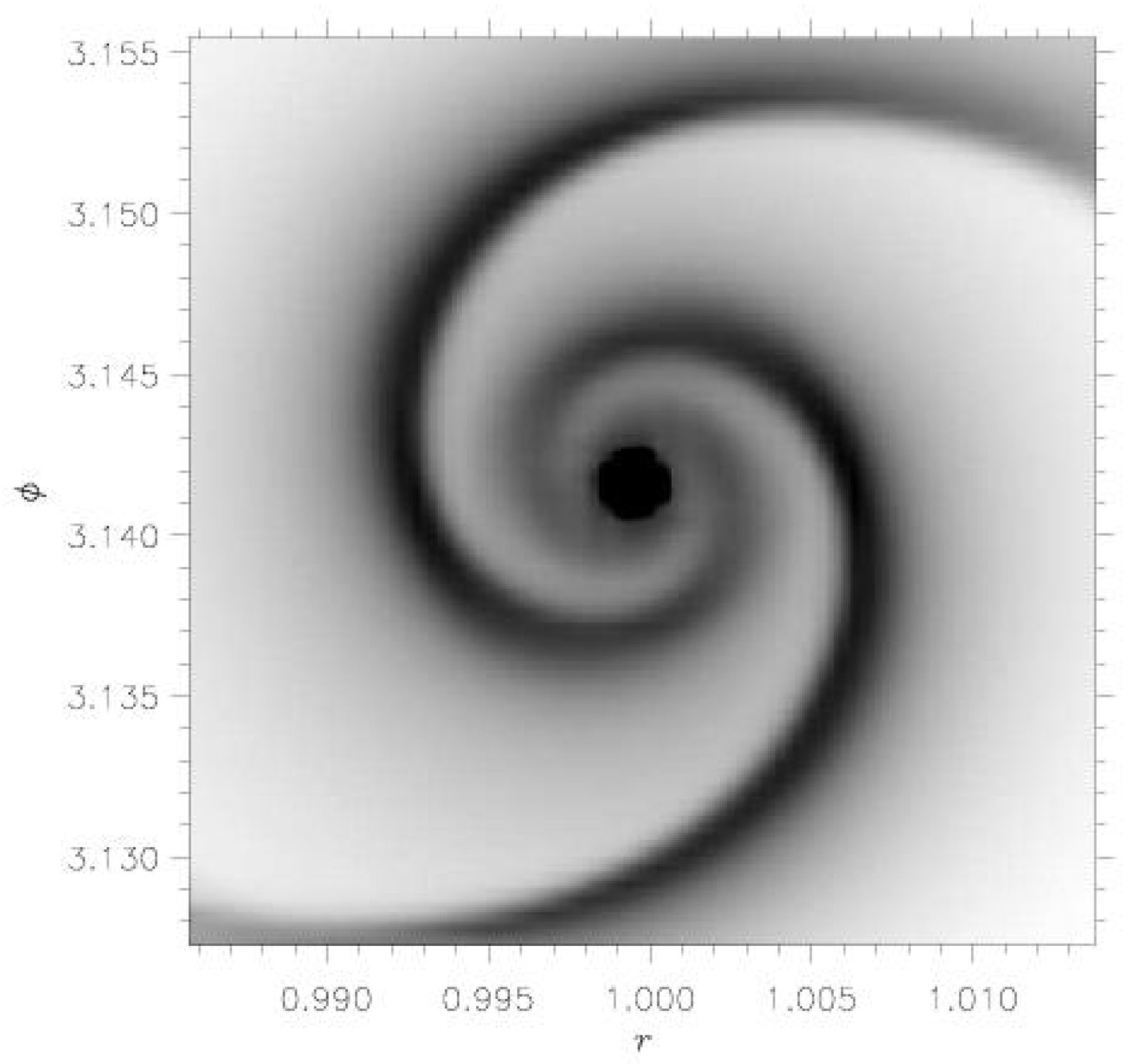}}
\caption{Density Structure of a $1 M_{Jup}$ on each level of the nested
grid system, consisting of 6 grid levels in total. The top left
panel displays the total computational domain. The line indicates the
size of the Roche lobe.
\label{fig:deep}}
\end{figure*}

\section{TYPE II MIGRATION}
\label{sec:typeII}
When the planet grows in mass  the disk response cannot be treated any
longer  as  a  linear  perturbation.  The  flow  perturbation  becomes
non-linear and the  planetary wake turns into a shock  in its vicinity. 
 Dissipation by these shocks as well as the action of viscosity leads to the deposition  of
angular momentum,  pushes material away from the planet and a gap opens.  The
equilibrium width of the gap is determined by the balance of
gap-closing viscous and pressure forces and gap-opening gravitational
torques. 

To obtain rough estimates, the condistion that the planet's gravity be strong enough
to overwhelm pressure in its neighbourhood is that the radius of the Hill sphere
exceed the disk semi-thickness or
\begin{equation} a\left({M_p \over 3 M_*}\right)^{1/3} > H. \label{Thc}
\end{equation}
The condition that angular momentum transport by viscous stresses
be interrupted by the planetary tide is approximately
\begin{equation} \left({M_p \over M_*}\right) > (40 a^2 \Omega)/\nu. \label{viscc}
\end{equation} 
For more discussion of these aspects of gap
opening,  see the review by
{\it Lin and Papaloizou} (1993), {\it Bryden et al.} (1999), and also
{\it Crida et al.}  (2006).

Interestingly for the standard parameters, $h =0.05,$
and $\nu/(a^2\Omega) = 10^{-5},$ (\ref{Thc}) gives
$M_p/M_* > 3.75\times 10^{-4},$ while
(\ref{viscc}) gives  $M_p/M_* >  4\times 10^{-4}.$
Note that this result obtained from simple estimates is in good
agreement with that obtained from Fig.~\ref{fig:mig1}.

Accordingly, for typical protoplanetary disk parameters, we can expect that a 
planet with a mass exceeding that of Saturn
will begin to open a visible gap. Using Eq. (\ref{eq:tanaka}) for the type~I
migration rate together with Eq. (\ref{Thc}), we can estimate the mimimum drift time at the marginal
gap opening mass to be given by
\begin{equation}
( a/\dot{a}) = 
3^{-2/3}(5.4+2.2\alpha)^{-1}\left({M_p\over M_*}\right)^{-1/3}\frac{M_*}{\pi\Sigma a^2}P_{orb},
\end{equation}
with $P_{orb}$  being the orbital period.
For a minimum mass solar nebula  with $ 4\pi\Sigma a^2= 2\times 10^{-3}M_*,$
and $M_p/M_* = 4\times 10^{-4}$ at 5.2~$au,$ this gives a minimum  drift time
of only  $\sim 3\times 10^4 y.$
This simply obtained estimate 
is in good agreement with the results presented in Fig.~{\ref{fig:mig1}.

\subsection{Numerical modelling}
Currently, heavy reliance is placed on numerical  methods  to analyse the
dynamics of the planet-disk  interaction, the density structure of the
disk, and  the resulting gravitational torques acting on  the planet.
The corresponding migration regime is called Type~II migration (e.g., {\it Lin and  Papaloizou}, 1986;
{\it Ward}, 1997).

The first modern  hydrodynamical calculations of
planet-disk interaction in the Type~II regime were performed by
{\it Bryden et al.} (1999), {\it Kley} (1999), {\it Lubow et al.} (1999). Since
protoplanetary accretion disks are assumed to be vertically thin,
these first simulations used a two-dimensional ($r-\varphi$) model of
the accretion disk.  The vertical thickness $H$ of the disk is
incorporated by assuming a given radial temperature profile $T(r)
\propto r^{-1}$ which makes the ratio $H/r$ constant. Typically the
simulations assume $H/r = 0.05$ so that at each
radius, $r,$  the Keplerian speed is 20 times faster than the local sound
speed.  Initial density profiles typically have power laws for the
surface density $\Sigma \propto r^{-s}$ with $s$ between $0.5$ and
$1.5$.  More recently,  fully 3D models have been calculated. These have  used
 the same kind of isothermal equation of state
({\it Bate et al.}, 2003; {\it D'Angelo et al.}, 2003a).

The  viscosity  is dealt with by solving the Navier Stokes
equations with
the kinematic viscosity $\nu$ taken as constant or given by an
$\alpha$-prescription $\nu = \alpha c_s H$, where $\alpha$ is a constant.
 From observations of protostellar disks,
 values lying between $10^{-4}$ and $10^{-2}$ are inferred for the
$\alpha$-parameter but there is great uncertainty.  Full MHD-calculations
have shown that the viscous stress-tensor ansatz may give (for
sufficiently long time averages) a reasonable approximation to the
{\it mean} flow in a turbulent disk ({\it Papaloizou and Nelson}, 2003).  The
embedded planets are assumed to be point masses (using a smoothed or softened
potential). The disk influences their orbits through 
gravitational torques which
 cause orbital evolution.  The planets may also
accrete mass from the surrounding disk ({\it Kley}, 1999).

\subsection{Viscous laminar Disks}
The type of modeling outlined in the previous section yields in
general smooth density and velocity profiles, and we refer to those
models as {\it viscous laminar disk} models, in contrast to models
which do not assume an a priori given viscosity and rather model the
turbulent flow directly (see below).

A typical result of such a viscous computation obtained with a $128
\times 280$ grid (in $r - \varphi$) is displayed in
Fig.~\ref{fig:pm3-200}. Here, the planet with mass $M_p = 1 M_{Jup}$
and semi-major axis $a_p = 5.2$AU is {\it not} allowed to move and
remains on a fixed circular orbit, an approximation which is made
in many simulations.  Clearly seen are the major effects an embedded
planet has on the structure of the protoplanetary accretion disk.  The
gravitational force of the planet leads to spiral wave patterns in the
disk. In the present calculation (Fig.~\ref{fig:pm3-200}) there are two
spirals in the outer disk and in the inner disk.  The tightness of the
spiral arms depends on the temperature (i.e. $h$) of the disk. The
smaller the temperature the tighter the spirals. The density gap at the location of the planet
discussed above is also visible. 

To obtain more insight into the flow near the planet and to
calculate accurately the torques of the disk acting on the planet,
the nested-grid approach described above has been used
together with a variable grid-size.
({\it D'Angelo et al.}, 2002; {\it Bate et al.}, 2003; {\it D'Angelo et al.}, 2003a).
Such a grid-system is fixed and therefore not adaptive.
The planet is located at the center of the finest grid.

The result for a 2D computation using 6 grids is displayed in
Fig.~\ref{fig:deep}, for more details see also
{\it D'Angelo et al.} (2002).  The top left base grid has a resolution
of $128 \times 440$ and each sub-grid has a size of $64 \times 64$
with a refinement factor of two from level to level.  It is noticeable
that the spiral arms inside the Roche-lobe of a high mass planet are detached
from the global outer spirals. The top right hand panel of figure~5 indicates
that the outer spirals fade away exterior to the one  around the planet.
 The two-armed spiral around the planet
extends deep inside the Roche-lobe and  enables the accretion of
material onto the planet.  The nested-grid calculations have recently
been extended to three dimensions (3D) and a whole range of planetary
masses has been investigated ({\it D'Angelo et al.}, 2003a).  In
the 3D case the spiral arms are weaker and accretion
occurs primarily from regions above and below the midplane of the
disk.
%
%

\subsection{The migration rate}
\label{subsubsec:migrat}
Such high-resolution numerical computations allow for a detailed
computation of the torque exerted by the disk material onto the
planet, and its mass accretion rate. 
For migration rates see  Fig.~\ref{fig:mig1}. 

The consequences of accretion and migration have been studied by
numerical computations which do not hold the planet fixed at some
radius but rather follow the orbital evolution of the planet
({\it Nelson et al.}, 2000), allowing planetary growth.  The typical
migration and accretion timescales are of the order of $10^5$~yrs,
while the accretion timescale may be slightly smaller.  This is in
very good agreement with the estimates obtained from the models using
a fixed planet.  These simulations show that during their inward
migration they grow up to about 4 $M_{Jup}$.

The inward migration time of $10^5$~yrs can be understood as the natural
viscous evolution time of the local accretion disk.
When the planet mass is not too large, and it makes a gap in the disk,
it tends to move as a disk gas particle would and thus move inwards
on the viscous timescale $\tau \sim a^2/\nu,$ ({\it Lin and  Papaloizou}, 1986).
But note that when the mass of the planet exceeds the disk mass in its
neighbourhood on a scale $a,$ the migration rate decreases because of
the relatively large inertia 
of the planet (see, e.g., {\it Syer and Clarke}, 1995; {\it Ivanov, Papaloizou and Polnarev}, 1999).

The consequence of the inclusion of thermodynamic effects (viscous
heating and radiative cooling) on the gap formation process
and type~II migration has been studied by
{\it D'Angelo et al.}, (2003b). In two-dimensional calculations  an
increased temperature in the circumplanetary disk has been found. This
has interesting consequences for the possible detection of an embedded
protoplanet. The effect that self-gravity
of the disk has on migration has been analysed through numerical
simulations ({\it Nelson and Benz}, 2003). For
typically expected protostellar  disk masses the influence is rather small.

\subsection{Consequences for evolution in young planetary systems}
A number of studies with the object of explaining the existence and distribution of giant planets
interior to the 'snow line' at $~2 au$ which make use of type~~II migration
have been performed
 (e.g., {\it Trilling et al.}, 1998; 2002; {\it Armitage et al.} 2002;  {\it Alibert et al.} 2004; {\it Ida and  Lin}, 2004). These assume formation
 beyond the snow line followed by inward type~II migration that
 is stopped by one of:  disk dispersal, Roche lobe overflow, stellar tides
 or entering a stellar magnetospheric cavity. Reasonable agreement
with observations is attained. But type~I migration has to suppressed
possibly by one of the mechanisms discussed in this review.
\section{TYPE III MIGRATION}
\label{sec:typeIII}
The terminology type~III migration  refers to migration for which an
important driver is material flowing through the coorbital region.
Consider an inwardly
(resp. outwardly) migrating  planet. Material of the inner disk
(resp. outer disk)  has to flow  across the  co-orbital region  and
it executes one U-turn  in the  horseshoe region (see Fig.~\ref{fig:horseshoe})
 to do so. By  doing this, it  exerts a
corotation torque  on the planet that  scales with the  migration rate. 
(see {\it Masset and Papaloizou}, 2003; {\it Artymowicz}, 2004; {\it Papaloizou}, 2005) for further analysis and discussion).

The specific angular momentum  that a  fluid element  near the
separatrix takes from  the planet when it switches  from an orbit with
radius $a-x_s$ to $a+x_s$ is $\Omega ax_s$
where  $x_s$  is  the   radial half  width  of the  horseshoe  region
estemated to be $2.5$ Hill sphere radii ({\it Artymowicz}, 2006).

The torque exerted on a planet migrating at a rate
$\dot a$ by the inner or outer disk elements as they cross the planet
orbit on a horseshoe U-turn is accordingly, to lowest order in $x_s/a$:
\begin{equation}
\Gamma_2 = (2\pi a\Sigma_s\dot a)\cdot (\Omega a x_s),
\end{equation}
where  $\Sigma_s$  is  the  surface  density at the  upstream
separatrix.  The system of  interest for
the evaluation of the  sum of external torques is
composed of the planet, all fluid elements trapped in libration in
 the   horseshoe region (with mass
$M_{HS}$) and the Roche lobe content (with mass $M_R$), because these
components migrate together.

\begin{figure}
\centering
\includegraphics[width=0.80\linewidth]{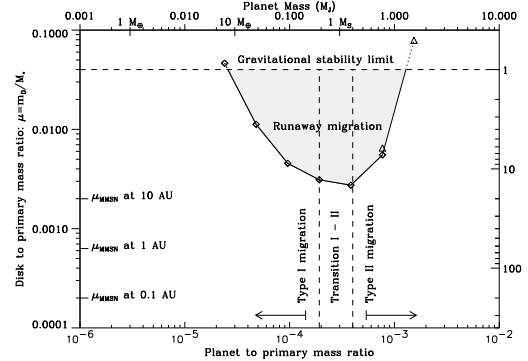}
\caption{\label{fig:rwlimit}
Runaway limit domain for a $H/R=0.04$ and $\nu=10^{-5}$ disk, with
a surface density profile $\Sigma\propto r^{-3/2}$. The variable
$m_D=\pi\Sigma r^2$ features on the $y$ axis. It is meant to represent
the local disk mass, and it therefore depends on the radius.
}
\end{figure}

\begin{figure}
\centering
\includegraphics[width=0.80\linewidth]{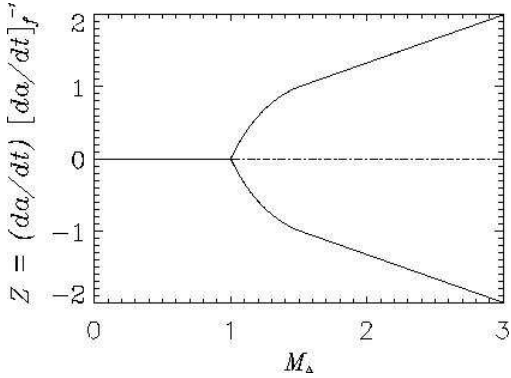}
\caption{\label{TY3}Bifurcation diagram derived from
equation (\ref{typethr}). This is extended to negative ${da\over dt}$
by making the curve symmetric about the $M_{\Delta}$ axis.
The bifurcation to fast migration occurs for $M_{\Delta } =1.$  }
\end{figure}

\begin{figure}
\centering
\includegraphics[width=0.80\linewidth]{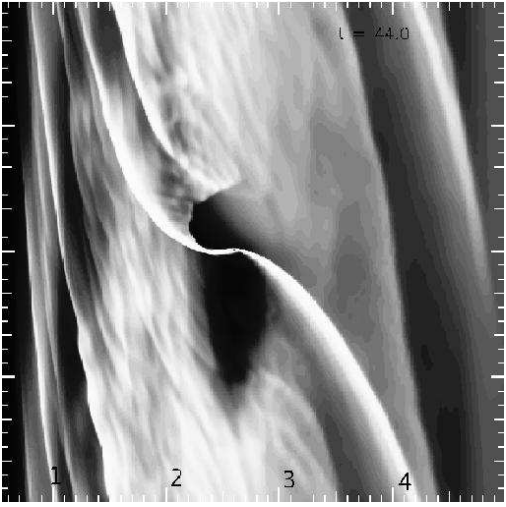}
\caption{\label{TY31}This is a density contour plot taken from a PPM
simulation with variable adaptive mesh
of a Jupiter mass planet in a disk  2.5 times more massive
than the minimum mass solar nebula (see {\it Artymowicz}, 2006).
         The planet after being placed on a positive density gradient,
increased its semi-major axis by a factor of 2.6
 in only 44 orbits. A trapped coorbital low density region
 is clearly visible and the  migration speed corresponds
to $M_{\Delta} \approx 3.$ }
\end{figure}

The drift rate of this system is then given by
\begin{equation}
{1\over 2}(M_p+M_{HS}+M_R)\cdot(\Omega \dot a)=(4\pi ax_s\Sigma_s)\cdot{1\over 2}(\Omega  a\dot a)+\Gamma_{LR}
\end{equation}
which can be rewritten as
\begin{equation}
\label{eqn:typeiiione}
M'_p\cdot {1\over 2}(\Omega \dot a)=(4\pi a\Sigma_sx_s-M_{HS})\cdot{1\over 2}
(\Omega a\dot a)+\Gamma_{LR},
\end{equation}
where $M'_p=M_p+M_R$ is all of the mass content within  the  Roche lobe, 
which  from now on for convenience  we refer to  as the  planet mass.
 The first term in the first bracket of the
r.h.s. of Eq~(\ref{eqn:typeiiione}) corresponds  to the horseshoe region surface  multiplied by the
upstream separatrix  surface density,  hence it is  the mass  that the
horseshoe region would have if it had a uniform surface density equal
to  the upstream  surface density.  The second  term  is the  actual
horseshoe  region mass.   The difference  between these  two  terms is
called in  {\it Masset and Papaloizou} (2003) the coorbital  mass deficit and
denoted $\delta m$. Thus we have
 
\begin{equation}
\label{eqn:typeiiitwo0}
  {1\over 2} \dot a\Omega a(M'_p-\delta m)=\Gamma_{LR}
\end{equation}
 Eq~(\ref{eqn:typeiiitwo0}) gives a drift rate
\begin{equation}
\label{eqn:typeiiitwo}
  \dot a=\frac{2\Gamma_{LR}}{\Omega a(M'_p-\delta m)}
\end{equation}
This  drift rate is  faster than  the standard  estimate in  which one
neglects  $\delta m$.   This comes  from the  fact that  the coorbital
dynamics alleviates the differential Lindblad torque task by advecting
fluid  elements from the  upstream to  the downstream  separatrix. The
angular momentum exchanged with the planet by  doing so produces a positive feedbak on its
migration.

As $\delta m$ tends to $M'_p$, most of the angular momentum change  of the
planet  and its  coorbital  region  is balanced  by  that of the orbit  crossing
circulating material, making migration increasingly cost effective.

When  $\delta  m\geq  M'_p$,  the  above analysis, assuming  a  steady
migration ($\dot a$ constant), is no longer valid. Migration may undergo
a  runaway leading to  a strongly time varying  migration rate. Runaway
(also denoted  type~III or fast ) migration  is therefore a  mode of  migration of
planets  that  deplete  their  coorbital  region and  are  embedded  in
sufficiently massive disks, so  that the above criterion be satisfied.
The  critical disk  mass above which a planet of given mass undergoes a
runaway depends on the disk parameters (aspect ratio and effective
viscosity). The limit has been considered by
{\it Masset and Papaloizou} (2003) for different disk aspect ratios and a
kinematic viscosity $\nu=10^{-5}$. 
The type III migration domain that was found 
for a disk with $H/r=0.04$ is indicated in figure~\ref{fig:rwlimit}.

Once the migration rate becomes fast enough that the planet migrates through
the coorbital region in less than a libration period, the analysis leading to
equation ~(\ref{eqn:typeiiitwo0}) becomes invalid. A recent analysis by 
{\it Artymowicz} (2006) indicates how this equation should be modified when this occurs.
Consider the torque term  $$\dot a\Omega a\delta m/2$$ in equation (\ref {eqn:typeiiitwo0})
above. This can be regarded as the corotation torque. It can be thought of as generated as follows.
Material enters the horseshoe region close to and behind the planet from the region into which the planet migrates at a maximum radial separation corresponding to the full half width $x_s.$
However, the turn at the oposite side of the horseshoe region $2\pi$ round in azimuth
occurs at a reduced maximum radial separation $x_s - \Delta$ due to the radial migration of the planet. Consideration of Keplerian circular orbits gives ({\it Artymowicz}, 2006)
\begin{equation}
\label{deltaf}
\Delta =   x_s\left(1 - \sqrt{ 1 - |\dot{a}|\dot{a}_f^{-1}}\right) \,,  
\end{equation}
where
\begin{equation}
  \frac{\dot{a}_f}{a} = \frac{3 x_s^2} {8\pi a^2}\Omega \  
\end{equation}
gives the critical or fast migration rate for which the horseshoe region can just extend the full $2\pi$
in azimuth. For larger drift rates it contracts into a tadpole like region
and the dynamics is no longer described by the above analysis.
Instead we should replace the square root in (\ref{deltaf}) by zero. 
For smaller drift rates, the torques exerted at  the  horseshoe turns on opposite
sides of the planet are proportional 
to $x_s^3$ and $(x_s -\Delta)^3 $ respectively. This is because these torques 
are proportional to the product of the flow rate,  specific angular mommentum transfered and the  radial width, each of them also being  proportional to the radial width.
As these torques act in opposite senses, the corotation torque should be proportional
to $x_s^3 - (x_s -\Delta)^3 ,$ or $x_s^3( 1- ( 1 - |\dot{a}|\dot{a}_f^{-1})^{3/2}).$
Note that this factor, which applies to both librating and non librating material, being $  (3/2)x_s^3|\dot{a}|\dot{a}_f^{-1}$ for small drift rates,
provides a match to equation (\ref{eqn:typeiiitwo0}) provided the ${\dot a}$ multiplyng
$\delta m$ is replaced by $(2/3)\dot{a}_f{\rm sign}(\dot{a})( 1- ( 1 - |\dot{a}|\dot{a}_f^{-1})^{3/2}).$

Equation (\ref{eqn:typeiiitwo0}) then becomes

\begin{equation}
  {1\over 2} \Omega a ( M'_p{\dot a}  -{2\over 3}\delta m\dot{a}_f{\rm sign}(\dot{a})( 1- ( 1 - |\dot{a}|\dot{a}_f^{-1})^{3/2}) )=\Gamma_{LR}. \label{typ3}
\end{equation}

Interestingly, when Lindblad torques are small,
 equation (\ref{typ3}) now allows for the existence of steady
fast migration rates which can be found by seting the left hand side to  zero.
Assuming without loss of generality that ${\dot a} >0,$ and setting
$Z= {\dot a}/{\dot a}_f,$ these states satisfy
\begin{equation}
\label{typethr}
Z= (2/3)M_{\Delta}(1 - (1-Z)^{3/2}),
\end{equation}
with $M_{\Delta} = (\delta m /M'_p).$
Equation (\ref{typethr}) gives rise to a bifurcation from the solution $Z=0$ to fast migration
solutions when $M_{\Delta}>1$ or when the coorbital mass deficit exceeds the planet mass
(see Figure~\ref{TY3}). The 'fast' rate, $Z=1,$ occurs when $M_{\Delta}=3/2.$
For larger $M_{\Delta},$ $Z=2M_{\Delta}/3.$

Fast migration, for the same disk profile and planet mass, can be
directed either outwards or inwards, depending on the initial
conditions.  This type of planetary  migration 
is found to depend on its migration history, the ``memory'' of this
history being stored in the way the horseshoe region is populated,
i.e.  in the preparation of the coorbital mass deficit. Note that
owing to the strong variation of the drift rate,
the horseshoe streamlines are not exactly closed, so that the
coorbital mass deficit can be lost and the runaway can stall. This has
been observed in some numerical simulations, whereas others show
sustained fast migration episodes for Saturn or Jovian mass planets
that can vary the semi-major axis by large
factors in less than $100$ orbits (e.g., see Figure~\ref{TY31}).
 To date, it is still unclear how long such episodes can last
for, and what are the conditions, if any, for them to stall or to
be sustained for a long period.

Because of the need to take account of complex coorbital flows in a partially
gap forming regime close to the planet, 
the problem of type~III migration is very numerically challenging
and therefore not unexpectedly  issues of adequate 
numerical resolution and convergence remain outstanding.
{\it D'Angelo et al.} (2005) have undertaken numerical
simulations of runaway migration using a nested grid system 
that can give high reolution within the Roche lobe,
but not elsewhere in the simulation, and found that the outcome, 
stated to be the suppression of type~III migration, 
 was highly dependent on the torque calculation
prescription (more precisely on  whether the Roche lobe material was
taken into account or not in this calculation) and on the mesh
resolution. One of the main subtleties of coorbital dynamics is to
properly take into account the inertia of all the material trapped
(even approximately) in libration with the planet, be it the horseshoe
or circumplanetary material. Including the Roche lobe material in the
torque calculation, while for other purposes is is assumed to be non self-gravitating, 
introduces a discrepancy between the inertial mass of the migrating
object (the point like object plus the Roche lobe content) and its active
gravitational mass (the mass of the point like object). This
unphysical discrepancy can be large and severely alter the migration
properties, especially at high resolution where the Roche lobe is
flooded by disk material in a manner than strongly depends on the
equation of state. On the other hand {\it Masset and Papaloizou} (2003)   consider the
Roche lobe content as a whole, referred to as the planet for simplicity,
assuming that for a given mass $M'_p$ of this object, there always
exists a point-like object with mass $M_p <M'_p$ such that
the whole Roche lobe content has mass $M'_p$. In this case, one needs
to exclude the Roche lobe content from the torque calculation (since
the forces originating from within the Roche lobe are not external forces),
and this way one naturally gets inertial and active gravitational masses of the
Roche lobe content (migrating object) that both amount to $M'_p$.

Another way of looking at this issue is to realize that from considerations
of angular momentum conservation, the angular momentum changes producing 
the migration can be evaluated at large distances from the Roche lobe.
Suppose that the material inside the Roche lobe had some asymmetric structure
that produced a torque on the point like mass $M_p.$ Then in order to sustain
this structure in any kind of steady state, external interaction would have to provide
an exactly counterbalancing torque that could be measured in material exterior to the Roche lobe
( for more details see the review by {\it Papaloizou and Terquem}, 2006).

In addition to the above issues, the effect of viscosity on the libration region
through its action on the specific vorticity profile
or the consequences of specific vorticity generation at  gap edges (e.g., {\it Koller et al.}, 2005)
has  yet to be considered.

{\em Potential consequences for forming planets}.
The Minimum Mass Solar Nebula (MMSN) 
is not massive enough to allow giant planets  to
experience runaway migration.
This  is most likely for planets with  masses comparable to that of Saturn
in disks that do not need to be more massive than a few times the Minimum Mass Solar Nebula.

Thus runaway migration should it occur makes the tendency for the migration rate
to be a maximum, in the mass range associated with the onset of gap
formation and type~II migration, more pronounced. 
  This may be related to the fact that most of the extrasolar planets known
as ``hot Jupiters'', with a semi-major axis $a < .06$~AU, happen to have
sub-Jovian masses. Assuming that type~I migration is suppressed
by, e.g., MHD turbulence (see  below) a forming protoplanet, as it passes through the domain
of  fast migration, would migrate very fast towards the central object
and at the same time it would accrete
gas from the nebula. If the protoplanet 
is able to  form a deep gap before it reaches the central regions of the disk,
or stellar magnetosphere, it
enters the slow, type~II migration regime, having at least about a
Jupiter mass. Otherwise, it reaches the central regions  still as a sub-Jovian object.

\section{TURBULENT PROTOPLANETARY DISKS}
\label{sec:turb}
The majority of calculations examining the interaction between protoplanets
and protoplanetary discs have assumed that the disc is laminar.
The mass accretion rates inferred from observations of young stars,
however, require an anomalous source of viscosity to operate
in these discs.
The most likely source of angular momentum transport
is magnetohydrodynamic turbulence generated
by the magnetorotational instability (MRI) ({\it Balbus and Hawley}, 1991).
Numerical simulations performed using both the local shearing box
approximation (see {\it Balbus and Hawley}, 1998 for a review)
and global cylindrical disc models (e.g., {\it Papaloizou and  Nelson}, 2003 and references therein) 
indicate
that the nonlinear outcome of the MRI is vigorous turbulence, and dynamo
action, whose associated stresses can account for the  observed
accretion rates inferred for T Tauri stars.

\begin{figure}[ht]
\centering \includegraphics[width=1.0\linewidth]{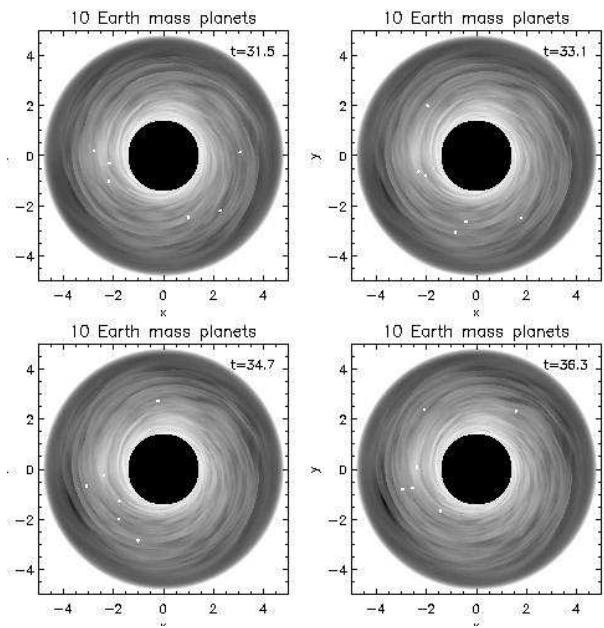}
\caption{\label{turb-fig1} This figure shows the orbital evolution of six 10 M$_{\oplus}$
protoplanets embedded in a turbulent protoplanetary disc.}
\end{figure}

These studies assumed that the approximation of ideal MHD 
was appropriate. The ionisation fraction in cool, dense protoplanetary discs,
however, is probably small in the planet forming region
between 1 -- 10 AU. Only the surface layers of
the disc, which are exposed to external sources of ionisation such
as X--rays from the central star or cosmic rays, are likely to be sufficiently
ionised to sustain MHD turbulence  
(e.g., {\it  Gammie}, 1996; {\it Fromang et al.}, 2002). However, this involves complex chemical reaction networks
and  the degree of depletion of dust grains which itself may vary while there is ongoing planet formation.

Recent work has examined the interaction between planets of various masses
and turbulent protoplanetary discs. These studies have usually 
simulated explicitly MHD turbulence arising from the MRI.
We now review the results of these studies.

\begin{figure}[ht]
\centering \includegraphics[width=1.03\linewidth]{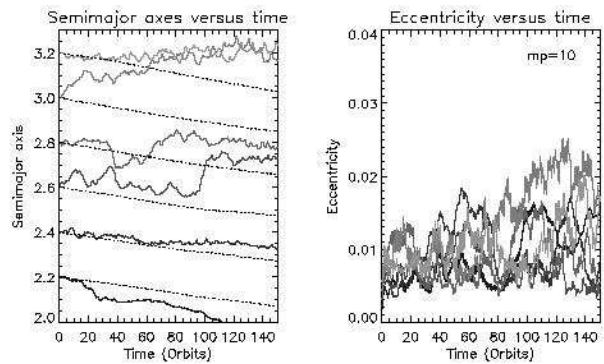}
\caption{\label{turb-fig2} The left panel shows the variation of semimajor
axis with time (measured in planet orbits at $r=2.3$). The dotted lines
represent the trajectories of 10 M$_{\oplus}$ planets in equivalent laminar 
discs, The right panel shows the variation of eccentricity. 
 For 10 M$_{\oplus}$ bodies, eccentricity damping due
to coorbital Lindblad torques  maintains low eccentricities.}
\end{figure}

\subsection{Low mass protoplanets in turbulent discs}
The interaction between low mass, non gap forming protoplanets and
turbulent discs has been examined by {\it Papaloizou et al.} (2004),
{\it Nelson and Papaloizou} (2004), {\it Nelson} (2005) and 
{\it Laughlin et al.} (2004). These calculations show that interaction
between embedded planets and density fluctuations generated by
MHD turbulence can significantly modify type I migration, at least over 
time scales equal to the duration of simulations that are currently 
feasible ($t\sim 150$ planet orbits).
leading to a process of `stochastic migration'
rather than the monotonic inward drift expected for planets
in laminar discs. Figure~\ref{turb-fig1} shows snapshots of the midplane 
density for six 10 M$_{\oplus}$ planets (non interacting)
embedded in a turbulent disc with $H/R =0.07,$
and show that the turbulent density fluctuations are of higher amplitude than
the spiral wakes generated by the planet 
({\it Nelson and Papaloizou}, 2004; {\it Nelson}, 2005).
Indeed, typical 
surface density fluctuations generated by turbulence in simulations
are typically $\delta \Sigma/\Sigma \simeq 0.15$ -- 0.3, with peak fluctuations
being ${\cal O}(1)$. Thus, on the scale of the disc thickness $H$, density
fluctuations can contain more than an Earth mass in a disc model 
a few times more massive than a minimum mass nebula.

Figure~\ref{turb-fig2} shows the variation in the semimajor axes of the
planets shown in figure~\ref{turb-fig1}, and figure~\ref{turb-fig3}
shows the running mean of the torque for one of the planets.
It is clear that the usual inward type I migration is
disrupted by the turbulence, and the mean torque does not converge toward
the value obtained in a laminar disc for the duration of the simulation.
A key question is whether
the stochastic torques can continue to overcome type I migration over
time scales up to disc life times. A definitive answer 
 will require very long global simulations.
\begin{figure}[ht]
\centering \includegraphics[width=0.8\linewidth]{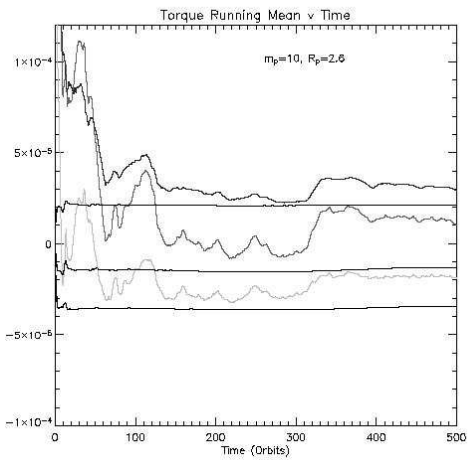}
\caption{\label{turb-fig3} This figure shows the running time average
of the torque per unit mass for the planet in figs.~\ref{turb-fig1}
and \ref{turb-fig2} whose initial orbit is at $r=2.6$.
The ultimately lowest but not straight line corresponds to the torque exerted by the
outer disc, the ultimately uppermost  line corresponds to the inner disc torque,
and the  line ultimately between them corresponds to the total torque.
The three  horizontal lines correspond to the inner, outer and
total torque exerted on a planet in laminar disc.}
\end{figure}

One can test the possibility that the response of the turbulent disc is
that of  a laminar disc 
whose underlying density is given by the time averaged value,
superposed on which are Gaussian distributed fluctuations
with a characteristic recurrence time $\sim$ the orbital period.
If such a picture applies then 
the time averaged torque experienced by the protoplanet, 
${\overline T}$, can be expressed as
\begin{equation}
{\overline T} = <T> + \frac{\sigma_T}{\sqrt{t_{tot}}}
\label{time-average}
\end{equation}
where $\sigma_T$ is the standard deviation of the torque amplitude,
$<T>$ is the underlying type I torque, 
and $t_{tot}$ is the total time elapsed, measured in units of
the characteristic time for the torque amplitude
to vary. Convergence toward the underlying type I value
is expected to begin  once the two terms on the right hand side become
equal.
For 10 M$_{\oplus}$ protoplanets, simulations indicate
that $\sigma_T \simeq 10 <T>$, with  a simple estimate of the time for
fluctuations to recur being $\simeq 1/2$ the planet orbital period 
({\it Nelson and Papaloizou}, 2004; {\it Nelson}, 2005). 
The torque convergence time is then $\simeq 50$ planet orbits.
Interestingly, the simulations presented in figure~\ref{turb-fig2}
were run for $\simeq 150$ planet orbits, and do not show a tendency for inward
migration.

Analysis of the stochastic torques suggests that they
vary on a range of characteristic times from the orbital period to
the run times of the simulations themselves ({\it Nelson}, 2005).
This feature can in principle allow
a planet to overcome type I migration for extended time periods.
It  appears to at least partially explain why the 
simulations do not show
inward migration on the time scale predicted by Eq.~\ref{time-average}.
The origin of these long time scale fluctuations is currently unknown.

Additionally, the picture described above of linear superposition
of stochastic fluctuations on an underlying type I torque may not
be correct due to non linear effects. Density fluctuations occuring within 
the disc in the planet vicinity are substantial, such that the usual bias
between inner and outer type I torques may not be recovered easily,
invalidating the assumptions leading to Eq.~\ref{time-average}.
To see this, consider a density fluctuation of order unity with length scale of
order $H$ a distance of order $H$ from the planet.  The characteristic
specific torque acting on it is $G\Sigma R.$ Given their  stochastic
nature, one might expect the specific torque acting on
the planet to oscillate between $\pm G\Sigma R.$ Note that this
exceeds the net specific torque implied by Eq.~(\ref{eq:tanaka}) by a
factor $T_f \sim [(M_{*}/M_p)(h)^3] h^{-1}.$ From the discussion in
section~\ref{sec:typeII}, the first factor should exceed unity for an
embedded planet. Thus, such an object is inevitably subject to large
torque fluctuations. The strength of the perturbation of the planet on
the disk is measured by the dimensionless quantity $( M_p/M_{*})(h)^{-3}.$ This perturbation
might be expected to produce a bias
in the underlying stochastic torques. If it produces
a non zero mean, corresponding to the typical fluctuation reduced
by a factor $T_f,$ this  becomes comparable to the laminar type~I
value. However, there is no reason to suppose the exact type~I result should be recovered.
But note the additional complication that the concept of such a mean may
not have much significance in practical cases, if large fluctuations can
occur on the disk evolutionary timescale,
such that it is effectively not established.

We comment that, as indicated in the plots in  figure~\ref{turb-fig3} at small times,
the one sided torque fluctuations can be more than an order of magnitude larger
than expected type I values. However, such fluctuations occur on an orbital timescale
and if averages over periods of $50$ orbits are considered values more like
type I values are obtained. Accordingly, the large fluctuations are not associated with
large orbital changes.

Further work is currently underway to clarify the role of turbulence
on modifying type I migration, including type I migration in
vertically stratified turbulent discs.

\subsection{High mass protoplanets in turbulent discs}
\begin{figure}[ht]
\centering \includegraphics[width=0.8\linewidth]{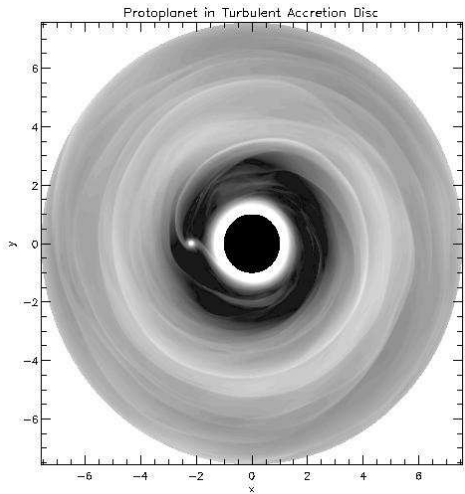}
\caption{\label{turb-fig4} This figure shows a snapshot of the disc
midplane density for a 5 M$_{\rm Jupiter}$ protoplanet embedded
in a turbulent disc.}
\end{figure}

The interaction of high mass, gap forming planets with
turbulent protoplanetary discs has been considered in a number of papers
({\it Nelson and Papaloizou}, 2003; {\it Winters et al.}, 2003;
{\it Papaloizou et al.}, 2004; {\it Nelson and Papaloizou}, 2004).
Figure~\ref{turb-fig4} shows
the midplane density for a turbulent disc with an embedded 5 M$_{\rm Jupiter}$ 
protoplanet. As expected from the discussion presented in section~\ref{sec:typeII},
gap formation occurs because the Roche lobe radius exceeds
the disc scale height, and tidal torques locally overwhelm viscous torques
in the disc.

{\it Papaloizou et al.} (2004) considered the transition
from fully embedded to gap forming planets using local shearing box
and global simulations of turbulent discs. These simulations showed
that gap formation begins when $(M_p/M_*)(R/H)^3 \simeq 1$,
which is the condition for the disc reponse to the planet gravity being non 
linear.
The viscous stress in simulations with zero--net magnetic
flux (as normally considered here) typically give rises to an effective 
$\alpha \simeq 5 \times 10^{-3}$, such that the viscous criterion for 
gap formation is satisfied when the criterion for non linear disc response 
is satisfied.

Global simulations allow the net torque on the planet due to the disc
to be calculated and hence the migration time to be estimated.
Simulations presented in {\it Nelson and
  Papaloizou} (2003, 2004) for
massive planets indicate migration times of $\sim 10^5$ yr, in line
with expections for type II migration.

\begin{figure}[ht]
\centering \includegraphics[width=0.8\linewidth]{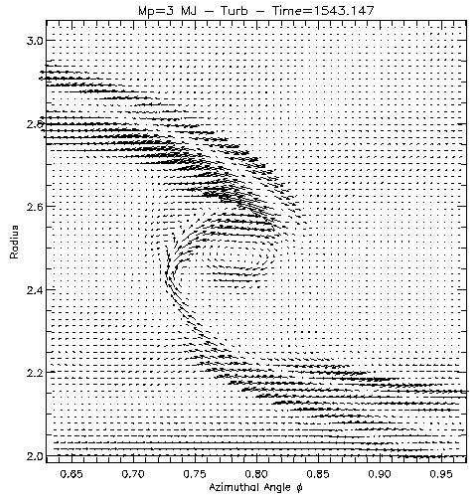}
\caption{\label{turb-fig5} This figure shows magnetic field lines
in the vicinity of the protoplanet. Field lines
link the protoplanetary disc with the circumplanetary disc 
within the planet Hill sphere.}
\end{figure}

\begin{figure}[ht]
\centering \includegraphics[width=0.8\linewidth]{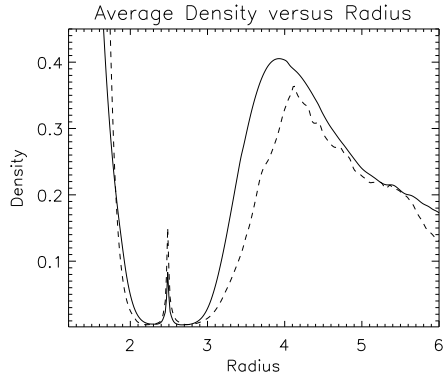}
\caption{\label{turb-fig6} The average density
in the vicinty of the planet as a function of radius for 
a laminar viscous disk (solid line) and for
a disk with   MHD turbulence (dashed line).
More mass has settled onto the planet in the latter case
possibly due to  magnetic braking.}
\end{figure}

A number of interesting features arise in simulations of gap forming
planets embedded in turbulent discs. The magnetic field topology
is significantly modified in the vicinity of the protoplanet, as
illustrated by figure~\ref{turb-fig5}.
The field is compressed and ordered in the postshock region associated
with the spiral wakes, increasing the magnetic stresses there.
Accretion of gas into the protoplanet Hill sphere causes advection of
field into the circumplanetary disc that forms there, and this field 
then links between the protoplanetary disc and the circumplanetary disc,
apparently contributing to magnetic braking of the circumplanetary material.
Indeed, comparison between a simulation of a 3 M$_{\rm Jupiter}$ 
protoplanet in a viscous laminar disc and an equivalent turbulent disc
simulation suggests that mass accretion onto the planet may be enhanced
by this effect. Figure~\ref{turb-fig6} shows the average density 
in the vicinity of the planet for a laminar disc run and a turbulent disc
simulation, indicating that the softened point mass used to model the
planet has accreted more gas in the turbulent disc run. In addition,
the gap generated tends to be deeper and wider in turbulent discs
than in equivalent viscous, laminar discs ({\it Nelson and Papaloizou}, 2003;
{\it Papaloizou et al.}, 2004). It is worth
noting, however, that the global simulations are of modest resolution,
and more high resolution work needs to be done to examine these
issues in greater detail.

\section{SUMMARY AND DISCUSSION} 
\label{sec:Disc}
 We  have reviewed recent progress in the field of disk planet interactions in the context
 of orbital migration. 
 This has mainly  come from large scale two and  three dimensional simulations that 
 have utilised the most
 up to date supercomputer resources. These have allowed the application of high resolution multigrid 
 methods and the study of disks with MHD turbulence interacting with planets.
 Simulations of both laminar and turbulent disks have been carried out and while
 the structure of the protoplanetary disk is uncertain, both these types are valuable.
 
 In the case of type~I migration it has become clear that the 3D simulations  discussed above agree with
 results obtained from linear analysis and  summing contributions from resonant torques 
  ({\it Tanaka et al.}, 2002)
 in predicting that planets in the $M_{\oplus}$ range in axisymmetric smoothly varying laminar disks,
 modelling the minimum mass solar nebula,
 undergo a robust inward migration on a $10^6$~$y$ timescale. This is a threat to the
 viability of the core accumulation scenario and a resolution has to be sought. 
 
 Several potential resolutions involving departures from the simple disk model
 have been suggested. These include invoking sharp radial opacity variations ({\it Menou and Goodman}, 2004),
  large scale nonaxisymmetric distortions
 (e.g., an eccentric disc) ({\it Papaloizou}, 2002) or a
  large scale toroidal magnetic field {\it Terquem}, (2003).
 
 If the disk is magnetically active, type~I drift may  be disrupted
 by stochastic migration  even for very extended periods of time ({\it Nelson and Papaloizou}, 2004).
  Even if this is unable to
 keep type~I migration at bay for the entire disk lifetime, the increased mobility of the cores
 may have important consequences for their build up through accretion 
 ({\it Rice and Armitage}, 2003).
 
 But note that even within the context of laminar disks, simulations have revealed that
 type~I migration has to be  suppressed in the $10M_{\oplus}$ range by weak nonlinear
 effects ({\it D'Angelo et al.}, 2003a).
 
 Gap formation in a standard MMSN disk is found to occur for planets with mass exceeding that of $\sim$Saturn,
 at which mass the migration rate is a maximum leading to an infall time of $~3\times 10^{4}$~$y.$
 Beyond that mass the rate decreases with the onset of type~II  migration on the viscous time scale.
 In more massive disks the maximum migration rate at a Saturn mass is potentially enhanced by
 the positive feedback from coorbital torques leading to a fast type~III migration regime
 ({\it Masset and Papaloizou}, 2003; {\it Artymowicz}, 2004). However the operation  of coorbital torques requires   to be clarified with further high resolution studies. But it is nonetheless interesting to
 speculate that the maximum migration rate at a Saturn mass is connected to
 the 'hot Jupiters' which tend to have sub Jovian masses (see section~\ref{sec:typeIII} above).
 Once  gap formation has occurred, type II migration ensues which appears to 
 allow  models that assume formation beyond the snow line to produce giant planet distributions
 in accord with observations (e.g., {\it Trilling et al.}, 1998; 2002; {\it Alibert et al.}, 2004).




\begin{thebibliography}{25}\setlength{\itemsep}{-2mm}


\bibitem[ali]{2004}  Alibert Y., Mordasini C., and Benz W. (2004)
 {\it  Astron. Astrophys.,  417},  L 25-L 28.
\item[] Armitage P. J.,  Livio M.,  Lubow S. H., and Pringle J. E. (2002)  
{ \it Mon. Not. R. Astr. Soc.,  334},  248-256.
\item[] Artymowicz P. (1993) {\it Astrophys. J., 419}, 155-165.
\item[] Artymowicz P. (2004) {\it Publ. Astr. Soc.  Pac., 324},  39-49.
\item[] Artymowicz P. (2006)  {\it In preparation}
\item[] Balbus S. A. and Hawley J. F. (1991) 
{\it Astrophys. J., 376}, 214-233.
\item[] Balbus S. A. and Hawley J. F,  (1998) {\it Rev. Mod. Phys.,  70}, 1-53.
\item[]
Balmforth N. J.  and Korycansky  D. G.  (2001)
 {\it Mon. Not. R. Astr. Soc.,  316},  833-851.
\item[] Bate M. R., Lubow S. H., Ogilvie G. I., and Miller K. A. (2003)
 {\it Mon. Not. R. Astron. Soc.,  341}, 213-229.
\item[] Bryden G., Chen X., Lin D. N. C., Nelson R. P., and Papaloizou J. C. B.
(1999) {\it Astrophys. J., 514}, 344-367.
\item[]
Crida A.,  Morbidelli A., and Masset F. S.,  (2006)  {\it Icarus, in press},  astro-ph/0511082.
\item[] D'Angelo G.,  Bate  M., and  Lubow S. (2005) 
{\it Mon. Not. R. Astr. Soc., 358}, 316-332.
\item[] D'Angelo G, Henning T., and Kley W. (2002) {\it Astron. Astrophys., 385}, 647-670.
\item[] D'Angelo G., Henning T., and Kley W. (2003a) 
{\it  Astrophys. J., 599}, 548-576.
\item[] D'Angelo G., Kley W., and Henning T. (2003b) 
{\it Astrophys. J., 586}, 540-561.
\item[] Fromang S., Terquem C., and Balbus S. A. (2002)
 {\it Mon. Not. R. Astron. Soc., 329 }, 18-28.
\item[]
Gammie C. F. (1996) {\it Astrophys. J., 457},  355-362.  
\item[]
Goldreich P. and  Sari R. (2003)  {\it Astrophys. J., 585}, 1024-1037.
\item[] Goldreich P. and Tremaine S. (1979) {\it Astrophys. J.,  233},
857-871.
\item[]
Ida S.  and  Lin D. N. C. (2004)  {\it  Astrophys. J.,  616},  567-572.  
\item[] Ivanov P. B., Papaloizou J. C. B., and Polnarev A. G. (1999) {\it
Mon. Not. R. Astron. Soc.,  307},  79-90.
\item[]Jang-Condell H., Sasselov D. D. (2005)
{\it Astrophys. J.,  619}, 1123-1131.
\item[] Kley W. (1999) {\it Mon. Not. R. Astron. Soc., 303}, 696-710.
\item[] Koller J., Li H. and Lin D. N. C. (2003) {\it Astrophys. J., 596},  L 91-L 94.
\item[] Korycansky D. G. and Pollack J. B. (1993)  {\it Icarus,  102}, 150-165.
\item[] Laughlin,  G., Steinacker A.,  and Adams F. C.  (2004)
 {\it Astrophys. J.,  608}, 489-496.
\item[] Lin D. N. C. and Papaloizou J. C. B. (1979) {\it
Mon. Not. R. Astron. Soc.,  186}, 799-830.
\item[] Lin D. N. C. and Papaloizou J. C. B. (1986) {\it Astrophys. J., 309}, 846-857.
\item[] Lin D. N. C.  and Papaloizou J. C. B.  (1993)
In  {\it Protostars and Planets III}  ( E. H. Levy and
J. I.  Lunine eds.),  pp. 749-835,
Univ. of Arizona, Tucson.
\item[] Lin D. N. C., Papaloizou J. C. B., Terquem C., Bryden G.,  and Ida S.
(2000) In {\it Protostars and Planets IV} ( V. Mannings et al. eds.), pp. 1111-1134, 
Univ. of Arizona, Tucson.
\item[] Lubow S. H., Seibert M., and Artymowicz P. (1999) {\it Astrophys. J., 526}, 1001-1012.
\item[] Marcy G. W. and Butler R. P. (1995) {\it 187th AAS Meeting BAAS., 27}, 1379-1384.
\item[] Marcy G. W. and Butler R. P. (1998) {\it Ann. Rev. Astron. Astr.,  36}, 57-97.
\item[] Mayor, M. and Queloz, D. (1995) {\it Nature, 378}, 355-359.
\item[] Masset  F. (2001)  {\it Astrophys. J., 558},  453-462.
\item[]
Masset F. (2002) {\it Astron. Astrophys.,  387},  605-623.
\item[]
Masset F. S., D'Angelo G.,  and  Kley W.  (2006)  {\it In preparation}.
\item[] Masset F. and Papaloizou J. C. B. (2003) {\it Astrophys. J., 588}, 494-508.
\item[] Mayor M. and Queloz D. (1995) {\it Nature, 378}, 355-359.
\item[]
Menou K.  and  Goodman  J. (2004) {\it Astrophys. J., 606}, 520-531.
\item[]
Nelson A. F. and Benz W. (2003) {\it Astrophys. J.,  589}, 578-604.
\item[] Nelson R. P. (2005) 
{\it Astron. Astrophys., 443}, 1067-1085.
\item[] Nelson R. P. and Papaloizou J. C. B. (2003)
 {\it Mon. Not. R. Astron. Soc., 339}, 993-1005.
\item[] Nelson R. P. and Papaloizou J. C. B. (2004)
 {\it Mon. Not. R. Astron. Soc., 350}, 849-864.
\item[] Nelson R. P., Papaloizou J. C. B., Masset F., and Kley W. (2000)
 {\it Mon. Not. R. Astron. Soc., 318}, 18-36.
\item[]
Ogilvie G. I. and  Lubow S. H.  (2003)  {\it Astrophys. J.,  587}, 398-406.
\item[] Papaloizou J. C. B. (2002) {\it Astron. Astrophys.,  388}, 615-631.
\item[] Papaloizou J. C. B.  (2005) {\it Celest. Mech. Dyn. Astron., 91}, 33-57.
\item[] Papaloizou J. C. B. and Larwood J. D. (2000)
 {\it Mon. Not. R. Astron. Soc., 315}, 823-833.
\item[] Papaloizou J. C. B. and Nelson R. P. (2003)
{\it Mon. Not. R. Astron. Soc., 350}, 983-992.
\item[] Papaloizou J. C. B. and Nelson R. P. (2005) {\it Astron. Astrophys., 433}, 247-265.
\item[] Papaloizou J. C. B., Nelson R. P., and Snellgrove M. D. (2004) 
{\it Mon. Not. R. Astron. Soc.,  350}, 829-848.
\item[] Papaloizou J. C. B.  and Terquem C. (2006) {\it
 Rep. Prog. Phys., 69}, 119-180.
\item[]
Pierens A. and Hur{\'e} J. M. (2005) {\it Astron. Astrophys., 433}, L 37-L 40.
\item[] Rice W. K. M. and  Armitage P. J. (2003) {\it Astrophys. J.,  598}, L 55-L 58.
\item[]  Syer D. and Clarke C. J. (1995) {\it Mon. Not. R. Astron. Soc.,  277}, 758-766.
\item[] Tanaka H., Takeuchi T., and Ward W. R. (2002) {\it Astrophys. J.,  565}, 1257-1274.
\item[] Terquem C. E. J. M. L. J. (2003) {\it Mon. Not. R. Astron. Soc., 341}, 1157-1173.
\item[] Trilling D. E., Benz W., Guillot T., Lunine J. I., Hubbard W. B., and
Burrows A. (1998) {\it Astrophys. J.,  500}, 428-439.
\item[] Trilling D. E., Lunine J. I., and Benz W. (2002) {\it Astron. Astrophys., 394}, 241-251.
\item[] Ward W. R.  (1997) {\it Icarus,  126}, 261-281.
\item[] Winters, W. F., Balbus S. A., and  Hawley, J. F.  
(2003)  {\it Astrophys. J.,  589},  543-555.

\end{thebibliography}
\end{document}